**On the nature of the spin frustration in the CuO$_2$ ribbon chains of LiCuVO$_4$: Crystal structure determination at 1.6 K, magnetic susceptibility analysis and density functional evaluation of the spin exchange constants**

Hyun-Joo Koo[1], Changhoon Lee[2], Myung-Hwan Whangbo[2,]*, Garry J. McIntyre[3§] and Reinhard K. Kremer[4,]*

[1] Department of Chemistry and Research Institute of Basic Science, Kyung Hee University, Seoul 130-701, Republic of Korea

[2] Department of Chemistry, North Carolina State University, Raleigh, North Carolina 27695-8204, USA

[3] Institut Laue-Langevin, BP 156, 38042 Grenoble Cedex 9, France

[4] Max-Planck-Institut für Festkörperforschung, Heisenbergstr. 1, D-70569 Stuttgart, Germany

[§] *Present address*: The Bragg Institute, Australian Nuclear Science and Technology Organization, Locked Bag 2001, Kirrawee DC NSW 2234, Australia

**Abstract**

The spin-1/2 $Cu^{2+}$ ions of $LiCuVO_4$ form one-dimensional chains along the b-direction, and the spin frustration in $LiCuVO_4$ is described in terms of the nearest-neighbor ferromagnetic exchange $J_1$ and the next-nearest-neighbor antiferromagnetic exchange $J_2$ in these chains. Recently, it has become controversial whether or not $J_1$ is stronger in magnitude than $J_2$. To resolve this controversy, we determined the crystal structure of $LiCuVO_4$ at 1.6 K by neutron diffraction, analyzed the magnetic susceptibility of $LiCuVO_4$ to deduce the Curie-Weiss temperature $\theta$ and the $J_2/J_1$ ratio, and finally extracted the spin exchange constants of $LiCuVO_4$ on the basis of density functional calculations. Our work shows unambiguously that the Curie-Weiss temperature $\theta$ of $LiCuVO_4$ is negative in the range of $-20$ K, so that $J_2$ is substantially stronger in magnitude than $J_1$.

## 1. Introduction

Lately much attention has been paid to the magnetic and dielectric properties of $LiCuVO_4$. It crystallizes with the inverse spinel structure and contains $CuO_2$ ribbon chains made up of edge-sharing $CuO_4$ square planes run along the crystallographic b-direction, and these chains are interlinked by corner-sharing $VO_4$ tetrahedra to form $CuVO_4$ layers parallel to the ab-plane. These are stacked along the c-direction with Li atoms occupying the sites between adjacent $CuVO_4$ layers (**Fig. 1a**). The $Cu^{2+}$ ($d^9$, S = 1/2) cations, the only magnetic ions in $LiCuVO_4$, form one-dimensional (1D) chains along the b-direction. A neutron diffraction study showed [1] that $LiCuVO_4$ exhibits an incommensurate antiferromagnetic (AFM) order below its Néel temperature $T_N \approx 2.4$ K. In this ordered magnetic structure, the spins lying in the $CuO_4$ planes have a spiral arrangement that propagates along the chain direction with the propagation vector q = (0, 0.532, 0). In general, a spiral spin order in a 1D chain of magnetic ions occurs when its nearest neighbor and next-nearest neighbor spin exchanges ($J_1$ and $J_2$, respectively) are spin frustrated.[2-4] Due to the loss of inversion symmetry associated with the spin spiral order, $LiCuVO_4$ becomes ferroelectric (FE) below $T_N$.[5-7] Density functional theory (DFT) calculations showed[8] that this FE polarization arises from the spin-orbit coupling interactions that occur in the absence of inversion symmetry.

In their inelastic neutron scattering study of $LiCuVO_4$, Enderle *et al.*[9] analyzed the spin-wave dispersion to deduce that $J_1$ is ferromagnetic (FM) ($J_1$ = 1.6 meV), the $J_2$ is AFM and is -3.6 meV (the bare exchange constant[9]) (**Fig. 1b**), so the $|J_2/J_1|$ ratio is substantially greater than 1 (i.e., $|J_2/J_1|$ = 2.3). They reproduced this observation by performing DFT calculations for $LiCuVO_4$ on the basis of its X-ray crystal structure (i.e., $|J_2/J_1| = 2.0$).[10] A recent inelastic neutron scattering investigation showing the two-spinon and four-spinon continuum [11] is also consistent with this observation. However, Sirker [12] re-analyzed the magnetic susceptibility and arrived at a strikingly different set of the spin exchanges (i.e., $J_1 \approx 7.8$ meV with $|J_2/J_1| \approx 0.5$). A similar conclusion was reported by Drechsler $et$ $al.$,[13] who employed the same Hamiltonian to fit the magnetic susceptibility and magnetization data of Enderle $et$ $al.$[9] Furthermore, their DFT calculations led to two sets of spin exchange constants for LiCuVO$_4$, which are drastically different from those they reported in ref. 9, namely, ($J_1 = 6.3$ meV, $J_2 = -5.1$ meV) and ($J_1 = 8.8$ meV, $J_2 = -6.5$ meV, $J_4 = -0.5$ meV, where $J_4$ refers to the inter-chain exchange $J_a$ defined in **Fig. 1b**).

The spin exchange constants appropriate for any magnetic solid should be consistent with its electronic structure, as evidenced for $(VO)_2P_2O_7$,[14,15] $Na_3Cu_2SbO_6$ and $Na_2Cu_2TeO_6$,[16-20] $Bi_4Cu_3V_2O_{14}$,[21-24] $Cu_3(CO_3)_2(OH)_2$,[25-27] and $Cu_3(P_2O_6OH)_2$,[28-30] to name a few. The magnetic structure of a given system is determined by its electronic structure, which depends critically on the accuracy of its crystal structure. In extracting the spin exchange constants of a magnetic solid in terms of electronic structure calculations, it is necessary that its crystal structure be accurate. In addition, the theoretical method of extracting spin exchange constants should be free of arbitrariness. In the method employed by Drechsler $et$ $al.$, one determines the electronic band structure of a magnetic solid by performing DFT calculations for its metallic state, then simulates the dispersion relations of the resulting partially-filled bands in terms of a set of hopping integrals, and finally converts these hopping integrals into the associated spin exchange constants. Using this "dispersion-simulation" method for LiCuVO$_4$, they reported $|J_2/J_1| \approx 2.0$ in ref.

9, but $|J_2/J_1| \approx 0.74$ and 0.81 in ref. 13. Spin exchange constants can be estimated more directly by the energy-difference mapping analysis[19,24,27,30,31] based on DFT calculations. In the latter method, one determines the relative energies for a set of ordered-spin magnetic insulating states by DFT calculations, and then equates their relative energies to the corresponding energies expected from the spin Hamiltonian defined in terms of the spin exchange parameters to determine (see below).

In the present work we attempt to resolve the aforementioned controversy concerning the relative magnitudes of $J_1$ and $J_2$ in LiCuVO$_4$. For a 1D magnetic chain defined by $J_1$ and $J_2$, the propagation vector q of its spin spiral is related to the $|J_2/J_1|$ ratio.[2-4] Unfortunately, for a spin-1/2 quantum system, this relationship is not sensitive enough to determine whether $|J_2/J_1| > 1$ or $|J_2/J_1| < 1$. In the mean-field approximation,[32] the Curie-Weiss temperature $\theta$ of a magnetic chain defined by $J_1$ and $J_2$ is given by $\theta \approx (J_1 + J_2)/2k_B$ (see below). This predicts that $\theta$ is negative if $|J_2/J_1| > 1$, but positive if $|J_2/J_1| < 1$. Consequently, provided that $\theta$ is accurately determined from the magnetic susceptibility of LiCuVO$_4$, one can decide which conclusion, $|J_2/J_1| > 1$ or $|J_2/J_1| < 1$, is correct. It should be noted that Enderle et al.[9,11] extracted the spin exchange constants for the crystal structure at 1.42 K,[9,11] but Drechsler et al.[13] employed the room-temperature crystal structure[10] for their calculations. In addition, the thermodynamic property study by Sirker[12] covered a wide range of temperature well above 1.42 K. Therefore, in resolving the controversy concerning the relative magnitudes of $J_1$ and $J_2$, it is necessary to check if the crystal structure of LiCuVO$_4$ undergoes any significant change when the temperature is lowered from room temperature.

Therefore, in the following, we first determine the crystal structure of $LiCuVO_4$ at 1.6 K by single-crystal neutron diffraction to ensure that the crystal structure of $LiCuVO_4$ does not undergo any significant change when the temperature is lowered. Then, we analyze the magnetic susceptibility of $LiCuVO_4$ in some detail to deduce its Curie-Weiss temperature $\theta$ and the $J_2/J_1$ ratio. When $\theta$ is small in magnitude, as is the case for $LiCuVO_4$, it is nontrivial to determine its sign unambiguously because $\theta$ is affected by other fitting parameters such as the g-factor and the temperature-independent contributions to the susceptibility (see below). Finally, we extract the spin exchange constants of $LiCuVO_4$ by performing the energy-difference mapping analysis based on DFT calculations. Our work shows unambiguously that the $|J_2/J_1|$ ratio is substantially greater than 1.

## 2. Single crystal structure at 1.6 K from neutron diffraction

Single crystals of $LiCuVO_4$ were grown from solutions of $LiCuVO_4$ in a $LiVO_3$ or $LiVO_3$-LiCl melt according to the procedures described in ref. 33. The composition and homogeneity of several crystals were checked by using electron microprobe energy dispersive X-ray analysis. The crystal selected for the present study (bar shaped size, $12 \times 4 \times 4$ mm$^3$) was identical to that used in several previous studies for the investigation of the magnetic structure and the spin wave excitations by elastic and inelastic neutron scattering.[1,9,11] Neutron diffraction performed on this crystal and a heat capacity measurement carried out on a piece (~13 mg) cut off from one end of the crystal indicated $T_N = 2.1(1)$ K.[1] Neutron diffraction was done at the ILL Grenoble on the diffractometer D10 with its unique four-circle dilution refrigerator to access low temperatures.[34] The neutron wavelength used in our study was λ = 2.354(2) Å, from a Cu 2 0 0 monochromated, and calibrated from measurements on a standard ruby crystal. The cell dimensions were refined using the ILL program RAFD9 and integrated intensities produced using the ILL program RACER.[35] The data were corrected for absorption in the crystal using the program DATAP.[36] Crystal structure refinements of the reduced squared structure factors ($F^2$) were performed using the program FULLPROF.[37]

The crystal was oriented at ambient temperature assuming the orthorhombic crystal structure derived in ref. 10. The sample was then quickly cooled to T = 1.6 K and a set of approximately 100 independent reflections was collected at this temperature. A full refinement of the nuclear structure was performed in the space group Imma (No. 74) by varying the positional parameters of the V, Cu and O atoms, the isotropic temperature factors of the V, Cu and O atoms, the scale factor and the extinction parameter. The Becker-Coppens Lorentzian model has been applied for the extinction correction and a mosaic spread was taken into consideration. The structure refinement was performed with least-squares methods on $F^2$. The results of these refinements are listed in **Table 1**. Such slight Li deficiency as found in the X-ray single crystal refinement [10] could not be detected in our neutron data. The atom positional parameters are in very good agreement with the room-temperature crystal structure data determined from X-ray diffraction previously.[10] A slight temperature induced lattice contraction is observed. There is no indication of a structural phase transition as $LiCuVO_4$ is cooled to 1.6 K. Thus, the temperature change does not cause any significant change in the crystal structure and hence the spin exchange constants of $LiCuVO_4$. In other words, the spin exchange constants deduced from the neutron scattering experiments are also appropriate for discussing other magnetic properties such as the magnetic susceptibility and magnetization data.

### 3. Analysis of magnetic susceptibility

The crystal used for the magnetic susceptibility measurement was different from the crystal used for the structure determination. It was flux-grown in a Pt-crucible from a mixture of Li-vanadate flux (see below) and CuO in the 1 : 0.4 ratio by slowly cooling the melt from 800 °C to 600 °C with a rate of 1 °C/h followed by a rapid cooling to room temperature. The flux was obtained by reacting a mixture of $Li_2CO_3$ and $V_2O_5$ in the 1.03 : 1 ratio (both materials purchased from Alfa Aesar, Puratronic with purity better than 99.99%) at 800 °C for ∼1 h. The crystals were mechanically separated from the solidified melt and remaining Li-vanadate flux on the crystals was washed off in ∼80 °C hot water.

A representative ensemble of crystals was selected and analyzed chemically by Inductively Coupled Plasma (ICP) Analysis (Labor Pascher, Remagen, Germany) for the Li, Cu and V content. The composition of the selected set of crystals was $Li_{1.003(11)}Cu_1V_{0.983(7)}O_4$ (batch E168). A well shaped crystal of ∼25 mg was selected and mounted with the $a$-axis along the magnetic field in the center of a Suprasil quartz tube and sealed under dried [4]He gas without using any glue in order to minimize addenda errors. The magnetic susceptibility was measured at 1 T with a SQUID magnetometer (MPMS-XL, Quantum Design).

The observed magnetic susceptibility data $\chi_{exp}(T)$ were fitted to the theoretical susceptibilities $\chi_{theory}(T)$ given by

$$\chi_{\text{theory}}(T) = \chi_{\text{spin}}(T) + \chi_0, \tag{1}$$

where the spin susceptibility, $\chi_{\text{spin}}$, can be discussed in two different ways. $\chi_{\text{spin}}$ can be described by the Curie-Weiss susceptibility

$$\chi_{\text{spin}} = \frac{N_A g_{\text{fit}}^2 \mu_B^2 S(S+1)}{3k_B(T-\theta)}, \tag{2}$$

where $N_A$ is Avogadro's number, $\mu_B$ the Bohr magneton, and $k_B$ the Boltzmann constant. Alternatively, $\chi_{\text{spin}}$ can be described by a high-temperature series expansion (HTSE) of the magnetic susceptibility for a frustrated S=1/2 Heisenberg chain defined by $J_1$ and $J_2 = \alpha J_1$,

$$H = -\sum_i \left( J_1 \hat{S}_i \cdot \hat{S}_{i+1} + J_2 \hat{S}_i \cdot \hat{S}_{i+2} \right) = -\sum_i \left( J_1 \hat{S}_i \cdot \hat{S}_{i+1} + \alpha J_1 \hat{S}_i \cdot \hat{S}_{i+2} \right). \tag{3}$$

For such a chain, the HTSE of the magnetic susceptibility is expressed as

$$\chi_{\text{spin}} = \frac{1}{T} \sum_{n,k} c_{n,k} \, \alpha^k \left( -J_1 / k_B T \right)^n, \tag{4}$$

where the expansion coefficients, $c_{n,k}$, were calculated by Bühler $et\ al.$[38] up to the $10^{\text{th}}$ order in $n$ and $k$.

At high temperatures where the spin susceptibility $\chi_{\text{spin}}$ is small in magnitude, the fitting analysis of the observed magnetic susceptibility is significantly influenced by the sign of the temperature-independent magnetic susceptibility $\chi_0$. An accurate estimation of $\chi_0$ is therefore essential for a meaningful fitting analysis of the susceptibility, for example, for a correct determination of the sign of the Curie-Weiss temperature $\theta$. $\chi_0$ consists of the diamagnetic contribution from the closed electron shell ions ($\chi_{\text{dia}}$) and the temperature-independent Van Vleck contribution ($\chi_{\text{VV}}$), i.e., $\chi_0 = \chi_{\text{dia}} + \chi_{\text{VV}}$. The diamagnetic contribution is well estimated from the incremental value for each atom in its respective oxidation state (Li$^+$:-0.6×10$^{-6}$ cm$^3$/mol; Cu$^{2+}$:-11×10$^{-6}$ cm$^3$/mol; V$^{5+}$:-4×10$^{-6}$ cm$^3$/mol; 4×O$^{2-}$: -12×10$^{-6}$ cm$^3$/mol).[39] Accordingly, $\chi_{\text{dia}} = -63.6 \times 10^{-6}$ cm$^3$/mol. The $\chi_{\text{VV}}$ of Cu$^{2+}$ depends on the direction of the applied magnetic field and can be estimated from the expression

$$(\chi_{\text{VV}})_c = 4(\chi_{\text{VV}})_a \approx 4(\chi_{\text{VV}})_b \approx 8 \frac{N_A \mu_B^2}{\Delta E}, \tag{5}$$

where $\Delta E$ is the energy separation from the singly-occupied x$^2$-y$^2$ orbital to the other occupied d orbitals. Here we assumed that the d orbitals other than the x$^2$-y$^2$ orbital are degenerate, which is a crude but reasonable approximation for the Cu$^{2+}$ ion in a square planar coordinate site. Then, $\Delta E \approx 2 - 3$ eV from optical spectroscopy data for typical Cu$^{2+}$ complexes.[40] Using these values for $\Delta E$ together with $N_A \mu_B^2 / k_B \approx 0.375$ cm$^3$K/mol, one obtains $(\chi_{\text{VV}})_a = (\chi_{\text{VV}})_b \approx +(20-32) \times 10^{-6}$ cm$^3$/mol. Similar values of $\chi_{\text{VV}}$ have been obtained, for example, for the Cu$^{2+}$ ions in YBa$_2$Cu$_3$O$_7$.[41] Consequently, $\chi_0 = \chi_{\text{dia}} + (\chi_{\text{VV}})_a \approx -(32-44) \times 10^{-6}$ cm$^3$/mol. Namely, it is most likely that $\chi_0$ is negative rather than positive.

In order to extract the values of g, θ and $\chi_0$ by fitting the modified Curie-Weiss law, Eq. (2), to the observed susceptibility data, we proceeded as follows:

(a) From the measured susceptibilities, we select a data set appropriate for the fitting analysis using the criteria that the selected temperature range should be sufficiently wide, the lower-boundary temperature of the selected temperature region should be sufficiently high so as to avoid the effect of short-range antiferromagnetic ordering, which makes the magnetic susceptibility deviate from a Curie-Weiss law. A closer inspection of our susceptibility data reveals slight deviations from a Curie-Weiss law towards the upper end of our dataset (above ~600 K). This is due probably to some experimental uncertainties such as the lowering of the sensitivity of the magnetometer and deterioration of the crystal at higher temperatures, an observation reported similarly by Krug *et al.*[42] The latter was found to occur between ~650 − 700 K for another crystal sample of the same batch. Thus, for the susceptibility fitting, we employ the susceptibility data in the temperature interval of 300 − 550 K.

(b) The initial least-square fits of the magnetic susceptibility by using g, θ and $\chi_0$ as free parameters showed a strong correlation between these parameters. Thus, in our fitting analysis, we treat only θ and $\chi_0$ as free parameters for a fixed g, and repeat this analysis for a series of g values from 2.03 to 2.16.

 **Fig. 2** displays the results of a representative fitting analysis using the g-factor of 2.07, which was found from an ESR experiment.[42] **Table 2** summarizes the results of the fitting analysis for various g values. For all g values employed (2.03 − 2.16), the Curie-Weiss temperature θ is negative. As g increases from 2.03 to 2.16, θ decreases gradually from -4 to -30 K while $\chi_0$ decreases gradually from a positive value ($37 \times 10^{-6}\,cm^3/mol$) to a negative value ($-21 \times 10^{-6}\,cm^3/mol$). Our estimate of $\chi_0$ based on $\chi_0 = \chi_{dia} + (\chi_{VV})_a$ indicated that $\chi_0$ is more likely to be negative than positive (see above). Therefore, **Table 2** suggests that θ is substantially negative ($-22$ K and lower). In addition, as judged from the $\chi^2$ values of the fitting analysis, the fitting is slightly better for the negative than for the positive $\chi_0$ values. Consequently, $|J_2/J_1| > 1$ in LiCuVO$_4$. This is in support of the conclusion from the neutron scattering studies.[9,11]

In the fitting analysis based on the Curie-Weiss law discussed above, we had to avoid the susceptibility data covering the temperature region where short-range AFM correlations occur. However, the HTSE fitting analysis is expected to be valid in the high temperature region where the Curie-Weiss law works as well as in the lower temperature region where short-range AFM correlations occur. Thus, we carried out the HTSE fitting analysis of the susceptibilities using Eq. (4) as follows:

(a) We chose the susceptibility data set for the HTSE fits covering $30 - 550$ K, which includes the onset of the short-range ordering anomaly centered at ∼27 K. Excluding the susceptibility data between $30 - 35$ K leads to a very slight change in the fitted parameters, without changing the general picture that the best fits are obtained with $J_1 = 10 - 12$ K and $\alpha = J_2/J_1 = -4 - -5$.

(b) We performed a series of least-squares fits to the selected dataset by using $\alpha$, $\chi_0$ and $J_1$ as free parameters for a fixed g value between 2.03 - 2.16.

**Fig. 3** displays the results of a representative HTSE fit of the experimental susceptibility data, and **Table 3** summarizes the results of the HTSE fitting analyses. The HTSE describes the high temperature susceptibilities well and is also able to capture the essence of the short range AFM ordering at lower temperatures. As already found from the Curie-Weiss fitting analysis, the Curie-Weiss temperature $\theta \approx (J_1 + J_2)/2k_B$ remains negative for all g values between $2.03 - 2.16$, and the best agreement with the experimental data is obtained for g $\approx 2.13$. The latter shows that $\alpha \approx -5$, $J_1 \approx 10$ K (i.e., 0.86 meV), and $\theta \approx -20$ K, which is consistent with the result of the Curie-Weiss fitting analysis (**Table 2**) and in good agreement with the Curie-Weiss temperature of $-15$ K obtained independently by Krug et al.[42] The fit of the HTSE provides a rather constant $J_2$, which is very close to that reported by Enderle et al.[9] and is independent of the g-factors chosen for the fitting.

## 4. Evaluation of spin exchange constants

For the spin exchange constants of LiCuVO$_4$, we consider the intra-chain exchanges $J_1$ and $J_2$ as well as the inter-chain exchanges $J_a$ and $J_{ad}$ (**Fig. 1**). (Here $J_a$ and $J_{ad}$ correspond to $J_4$ and $J_5$, respectively, in the notations of Enderle et al.[9]) The spin exchanges between adjacent CuVO$_4$ layers are not considered because they were found to be very weak both experimentally[9,11] and theoretically.[9,13] To evaluate the four spin exchange constants $J_1$, $J_2$, $J_a$ and $J_{ad}$, we carry out DFT calculations for the five ordered spin states (**Fig. 4**) defined by using the (2a, 2b, 2c) supercell of the 1.6 K crystal structure of LiCuVO$_4$. Our calculations employed the frozen-core projector augmented wave method encoded in the Vienna ab initio simulation package (VASP)[43] with the generalized-gradient approximation (GGA)[44] for the exchange-correlation functional, the plane-wave cut-off energy of 400 eV, and a set of 4×4×2 k-points for the irreducible Brillouin zone. To describe the possible effect of the strong electron correlation in the Cu 3d states, the GGA plus on-site repulsion U (GGA+U) method[45] was employed with effective $U_{eff}$ = 4, 5 and 6 eV on the Cu atom. The threshold for the self-consistent-field convergence of the total electronic energy was $10^{-6}$ eV.

For the five ordered magnetic states of LiCuVO$_4$, our spin-polarized GGA calculations with VASP show that they have a band gap at the Fermi level, in agreement with the fact that LiCuVO$_4$ is a magnetic insulator. The relative energies of the five ordered spin states obtained from the GGA+U calculations are summarized in **Fig. 4**. To extract the values of J$_1$, J$_2$, J$_a$ and J$_{ad}$ from these relative energies, we express the total spin exchange interaction energies of the five ordered spin states using the spin Hamiltonian defined in terms of J$_1$, J$_2$, J$_a$ and J$_{ad}$,

$$\hat{H} = -J_{ij}\hat{S}_i \cdot \hat{S}_j, \qquad\qquad\qquad (6)$$

where $\hat{S}_i$ and $\hat{S}_j$ are the spin operators at the spin sites i and j, respectively, and J$_{ij}$ (= J$_1$, J$_2$, J$_a$, J$_{ad}$) is the spin exchange parameter between the sites i and j. By applying the energy expressions obtained for spin dimers with N unpaired spins per spin site (in the present case, N = 1),[46] the total spin exchange energies, per two formula units (FUs), for the five spin states of LiCuVO$_4$ are written as

$$E_{FM} = (-2J_1 - 2J_2 - 2J_a - 4J_{ad})(N^2/4)$$

$$E_{AF1} = (-2J_1 - 2J_2 + 2J_a + 4J_{ad})(N^2/4)$$

$$E_{AF2} = (+2J_2 + 2J_a)(N^2/4)$$

$$E_{AF3} = (+2J_2 - 2J_a)(N^2/4)$$

$$E_{AF4} = (+2J_1 - 2J_2 - 2J_a + 4J_{ad})(N^2/4) \qquad\qquad (7)$$

Thus, by mapping the energy differences between the five ordered states determined from the GGA+U calculations onto the corresponding energy differences determined from the spin Hamiltonian, we obtain the values of the four spin exchange constants summarized in **Table 4**.

It is seen from **Table 4** that J$_1$ is FM, J$_2$ is AFM, and the |J$_2$/J$_1$| ratio is much greater than 1; |J$_2$/J$_1$| = 8.7, 5.6 and 4.1 using the J$_1$ and J$_2$ values obtained from the

GGA+U calculations with $U_{eff}$ = 4, 5 and 6 eV, respectively. In terms of the $|J_2/J_1|$ ratio, the $J_1$ and $J_2$ values from $U_{eff}$ = 6 eV are in best agreement with those deduced from the neutron scattering experiments.[9,11]

In the mean-field approximation,[32] the Curie-Weiss temperature θ is related to spin exchange parameters as

$$\theta = \frac{S(S+1)}{3k_B}\sum_i z_i J_i \approx \frac{(J_1 + J_2 + J_a + 2J_{ad})}{2k_B} \qquad (8)$$

where the summation runs over all nearest neighbors of a given spin site, $z_i$ is the number of nearest neighbors connected by the spin exchange parameter $J_i$, and S is the spin quantum number of each spin site (i.e., S = 1/2 in the present case). Therefore, the θ value is estimated to be −86, −66 and −50 K by using the spin exchange constants calculated from the use of $U_{eff}$ = 4, 5 and 6 eV, respectively. Given the general observation that GGA+U calculations overestimate the spin exchange constants by a factor of up to ~4,[19,24,47] the θ value for $LiCuVO_4$ is expected to be lower than −22, −17 and −13 K from the GGA+U calculations with $U_{eff}$ = 4, 5 and 6 eV, respectively. Since the calculations with $U_{eff}$ = 6 eV give the $|J_2/J_1|$ ratio in best agreement with the neutron scattering studies, the best estimate for θ is lower than −13 K from our calculations. The latter is in good agreement with the θ values deduced from the magnetic susceptibility analyses discussed in the previous section.

Provided the calculated spin exchange constants are overestimated by a factor of ~4, the calculated $J_2$ values of **Table 4** in good agreement with the values deduced from the magnetic susceptibility analysis (**Table 3**) and also with the bare spin exchange constant derived from the neutron scattering experiments.[9] As for the inter-chain spin exchanges, our calculations show that $J_a$ is FM, $J_{ad}$ is AFM, and $J_a$ is stronger than $J_{ad}$ (**Table 4**). These results are not in agreement with those extracted from the neutron scattering experiments (i.e., $J_a$ is AFM, $J_{ad}$ is FM, and $J_a$ is weaker than $J_{ad}$).[9] Nevertheless, our calculations are consistent with the experiments in that the dominant inter-chain spin exchange is FM rather than AFM.

## 5. Concluding remarks

The crystal structure of LiCuVO$_4$ determined by neutron diffraction at 1.6 K is very close to that determined by X-ray diffraction at room temperature except for a slight contraction in the cell parameters. Indication for a structural phase transition was not observed. Therefore, the spin exchange constants deduced from the neutron scattering experiments at 1.42 K are also relevant for discussing the temperature dependence of the magnetic susceptibility. Our analyses of the magnetic susceptibility of LiCuVO$_4$ show that the Curie-Weiss temperature θ is negative in the range of −20 K, and that the $|J_2/J_1|$ ratio should be substantially greater than 1. This observation is corroborated by the spin exchange constants of LiCuVO$_4$, which we evaluated from the energy-difference mapping analysis based on DFT calculations for the 1.6 K structure of LiCuVO$_4$. Thus, our work supports the conclusion by Enderle *et al*.[9,11] that the $|J_2/J_1|$ ratio is substantially greater than 1. The inter-chain spin exchange interactions of LiCuVO$_4$ are not negligible according to the spin exchange constants obtained from our energy-difference mapping analysis and also from those deduced from the neutron scattering experiments.

**Acknowledgments**

We thank E. Brücher, B. J. Gibson, and G. Siegle for experimental assistance, and A. V. Prokofiev and W. Aßmus for providing the crystal for the neutron structure determination. We are grateful to G. S. Uhrig and M. Enderle for invaluable comments. The work at North Carolina State University was supported by the Office of Basic Energy Sciences, Division of Materials Sciences, U. S. Department of Energy, under Grant DE-FG02-86ER45259.

Table 1.        Crystal structure of LiCuVO$_4$ determined by neutron diffraction at T = 1.6 K [Space group = Imma; a = 5.6477 (9) Å, b = 5.7864 (9) Å, c = 8.6940 (13) Å; R$_F^2$ = 9.2%, R$_F$ = 7.1%, $\chi^2$ = 2.1]. The isotropic temperature factors of Cu and V were treated as a single parameter to refine.

| Atom | Wycoff | x | y | Z | B$_{eq}$ (Å$^2$) |
|------|--------|---|---|---|------------------|
| Li | 4d | 0.25 | 0.25 | 0.75 | 0.3(0) |
| Cu | 4a | 0 | 0 | 0 | 0.10(1) |
| V | 4e | 0 | 0.25 | 0.3915 (36) | 0.10 (1) |
| O(1) | 8h | 0 | 0.0155(4) | 0.2742 (5) | 0.16 (15) |
| O(2) | 8i | 0.2359 (6) | 0.25 | -0.0012 (2) | 0.16 (15) |

Table 2.        Values of the g-factor g, the Curie-Weiss temperature θ, the temperature-independent susceptibility $\chi_0$ deduced from the fitting analysis by using the observed magnetic susceptibility data between $300 - 550$ K. The experimental susceptibility data employed are identical to those published in ref. 9.

| g | θ (K) | $\chi_0$ ($10^{-6}$ cm$^3$/mol) | $10^3 \times \chi^{2\,(a)}$ |
|---|---|---|---|
| 2.03 | -3.7(3) | +37.3(7) | 5.59 |
| 2.04 | -5.7(3) | +32.8(7) | 5.27 |
| 2.05 | -7.7(3) | +28.3(7) | 4.90 |
| 2.06 | -9.8(3) | +23.9(6) | 4.62 |
| 2.07 | -11.8(3) | +19.4(6) | 4.37 |
| 2.08 | -13.7(3) | +14.9(6) | 4.15 |
| 2.09 | -15.8(3) | +10.4(6) | 3.97 |
| 2.10 | -17.9(3) | +5.9(6) | 3.81 |
| 2.11 | -19.9(3) | +1.4(6) | 3.69 |
| 2.12 | -21.9(3) | -3.1(6) | 3.59 |
| 2.13 | -24.0(3) | -7.6(6) | 3.53 |
| 2.14 | -26.0(3) | -12.1(6) | 3.49 |
| 2.15 | -28.0(3) | -16.6(6) | 3.47 |
| 2.16 | -30.0(3) | -21.1(6) | 3.49 |

(a) $\chi^2$ is defined as $[1/(N-P)]\sum_i w_i [\chi_{exp}(T_i) - \chi_{theory}(T_i, g, \theta, \chi_0)]^2$, where $\chi_{exp}(T_i)$ and $\chi_{theory}(T_i)$ are the experimental and calculated data points at temperature $T_i$, respectively, and $w_i = 1/\chi_{exp}(T_i)$. P is the number of free parameters chosen to fit the experimental data (in our case, P = 2), and N is the number of data points used in the fit (in our case, N = 88). Since N and P have been kept identical in the fitting procedure, $\chi^2$ immediately reflects the quality of agreement between experiment and theory; the smaller $\chi^2$, the better the agreement.

Table 3. Values of g, $\alpha = J_2/J_1$, $\chi_0$ and $J_1$ obtained from the HTSE fitting analysis of the magnetic susceptibility data between $30 - 550$ K. The experimental susceptibility data employed are identical to those published in ref. 9. The fitted values are given with error bars in parentheses.

| g | $\alpha = J_2/J_1$ | $\chi_0$ ($10^{-6}$ cm$^3$/mol) | $J_1/k_B$ (K) | $J_2/k_B$ (K) | $\chi^2$ [a] |
|---|---|---|---|---|---|
| 2.03 | -3.4(6) | +67.7(9) | 12.7(8) | -43.2 | 0.327 |
| 2.04 | -3.5(6) | +59.5(9) | 12.5(8) | -43.8 | 0.274 |
| 2.05 | -3.5(5) | +51.3(9) | 12.4(7) | -43.4 | 0.226 |
| 2.06 | -3.7(5) | +43.2(8) | 12.0(7) | -44.4 | 0.184 |
| 2.07 | -3.8(4) | +34.9(8) | 11.8(7) | -46.0 | 0.146 |
| 2.08 | -4.0(3) | +26.7(8) | 11.5(6) | -46.0 | 0.114 |
| 2.09 | -4.1(3) | +18.5(8) | 11.3(5) | -46.3 | 0.086 |
| 2.10 | -4.2(3) | +10.2(7) | 11.0(5) | -46.2 | 0.064 |
| 2.11 | -4.4(2) | +2.0(7) | 10.6(5) | -46.6 | 0.047 |
| 2.12 | -4.7(2) | -6.2(7) | 10.2(4) | -48.0 | 0.035 |
| 2.13 | -5.1(2) | -14.3(7) | 9.5(4) | -48.5 | 0.027 |
| 2.14 | -5.8(3) | -22.2(7) | 8.3(4) | -48.1 | 0.025 |
| 2.15 | -7.8(4) | -29.4(8) | 6.2(4) | -48.4 | 0.027 |
| 2.16 | -15.5(6) | -35.9(9) | 3.0(6) | -46.5 | 0.031 |

(a) $\chi^2$ is defined as in Table 2 with P = 3 and N = 176.

Table 4. Spin exchange constants $J_1/k_B$, $J_2/k_B$, $J_a/k_B$ and $J_{ad}/k_B$ (in K) extracted from GGA+U calculations with $U_{eff}$ = 4, 5 and 6 eV and those deduced from the neutron scattering experiments by fitting a classical spin wave theory to the measured dispersion.

|            | $U_{eff}$ = 4.0 eV | $U_{eff}$ = 5.0 eV | $U_{eff}$ = 6.0 eV | ref. 9 |
|------------|--------|--------|--------|--------|
| $J_1/k_B$  | 25.0   | 31.3   | 34.9   | 18.6   |
| $J_2/k_B$  | -208.7 | -173.8 | -144.0 | -41.4  |
| $J_a/k_B$  | 17.6   | 15.2   | 13.1   | -0.1   |
| $J_{ad}/k_B$ | -2.0 | -1.9   | -1.9   | 4.6    |

**Figure captions**

Figure 1. (a) Perspective view of the crystal structure of $LiCuVO_4$, where the blue, cyan, red and yellow circles represent the Cu, V, O and Li, atoms, respectively. (b) Definitions of the intra- and inter-chain spin exchange paths, where the labels 1, 2, a and ad refer to the spin exchanges $J_1$, $J_2$, $J_a$ and $J_{ad}$, respectively.

Figure 2. Results of a representative fitting analysis for the magnetic susceptibility of $LiCuVO_4$ with a modified Curie-Weiss law (red solid line) using the g-factor of 2.07 and the experimental magnetic susceptibility data in the range of $300 - 550$ K with the parameters given in the inset.

Figure 3. Results of a representative HTSE fitting analysis (red solid line) for the magnetic susceptibility of $LiCuVO_4$ using a frustrated chain defined by $J_1$ and $J_2$ (= $\alpha J_1$) with g = 2.12. The experimental magnetic susceptibility data in the range of $30 - 550$ K were used for the fitting with the parameters given in the inset.

Figure 4. Five ordered spin states of $LiCuVO_4$ employed to extract the spin exchanges $J_1$, $J_2$, $J_a$ and $J_{ad}$ by using a (2a, 2b, c) supercell. For simplicity, only the $Cu^{2+}$ ions are shown, and the unshaded and shaded circles represent the $Cu^{2+}$ ions with up-spin and down-spin, respectively. The three numbers in the parenthesis (from left to right) for each state are the relative energies (in meV per 2 FUs) obtained from the GGA+U calculations with $U_{eff}$ = 4, 5 and 6 eV, respectively).

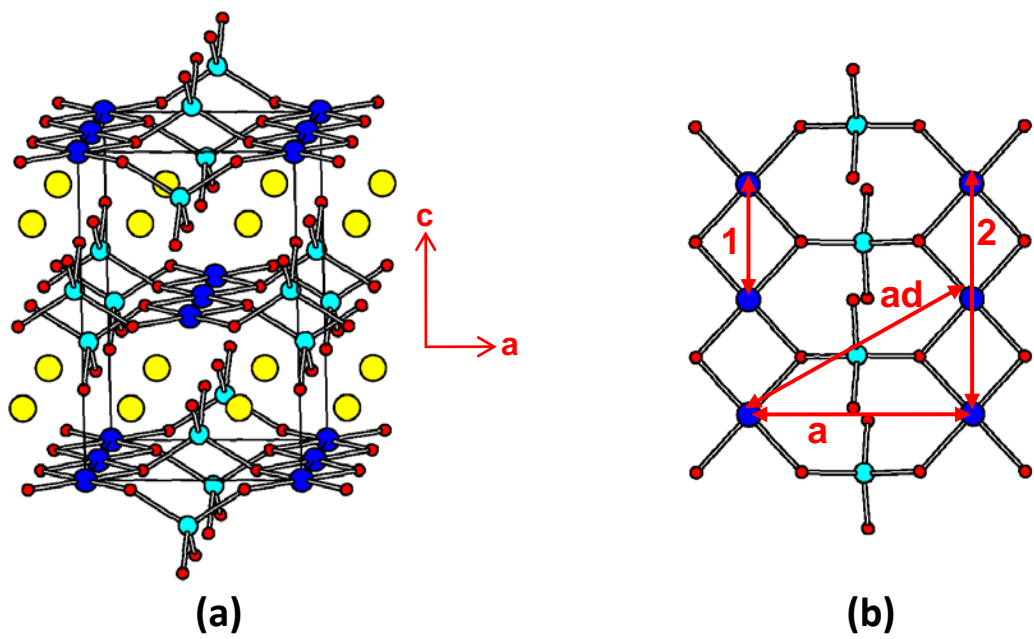

**(a)**                    **(b)**

Figure 1.

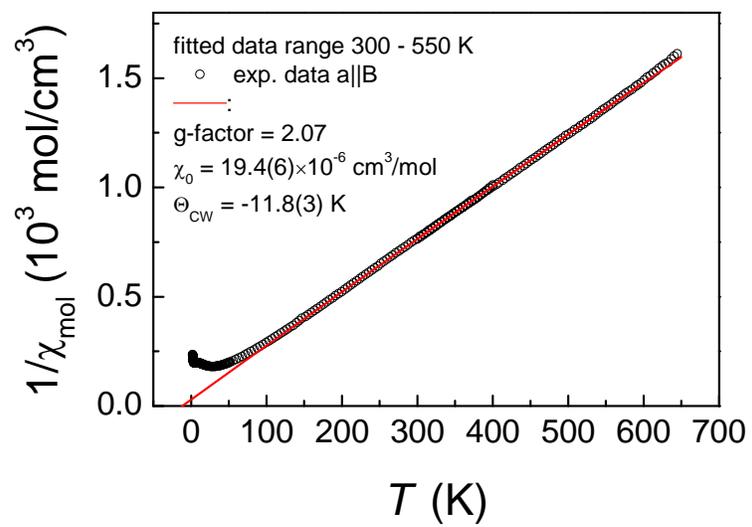

Figure 2.

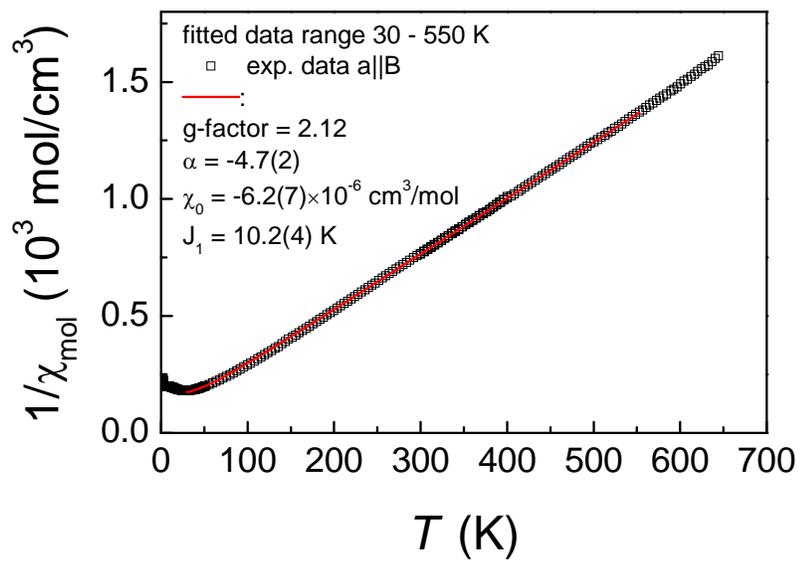

Figure 3

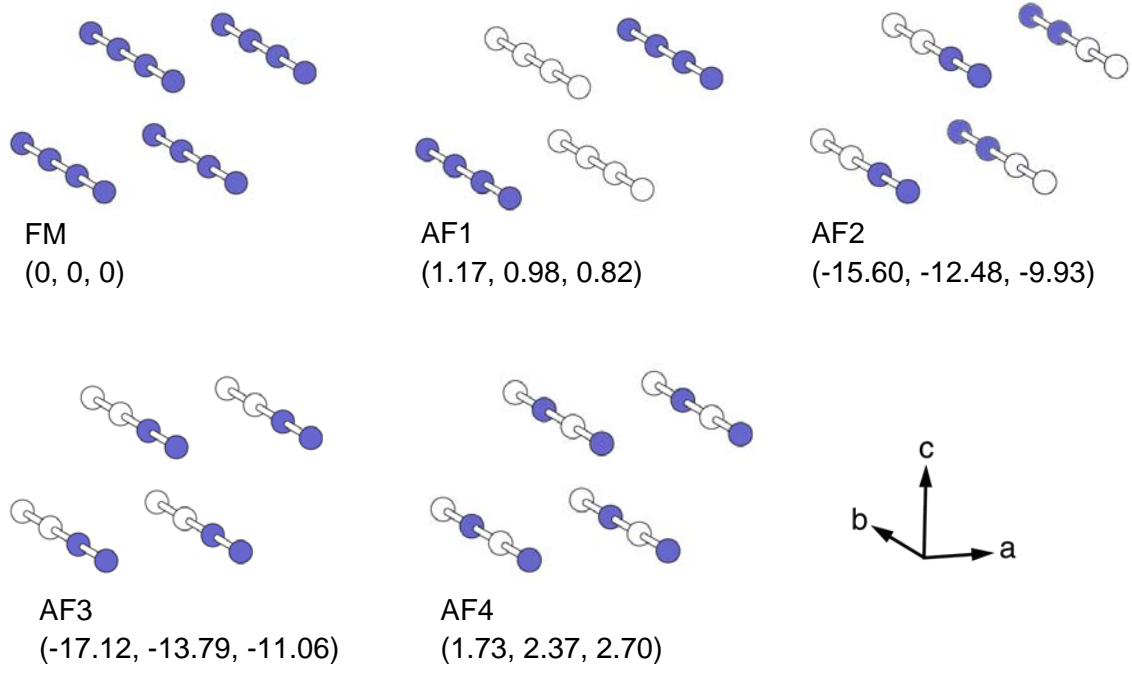

FM
(0, 0, 0)

AF1
(1.17, 0.98, 0.82)

AF2
(-15.60, -12.48, -9.93)

AF3
(-17.12, -13.79, -11.06)

AF4
(1.73, 2.37, 2.70)

Figure 4.

**Synopsis**

The spin frustration in the chains of $Cu^{2+}$ ions in $LiCuVO_4$ is described by the nearest-neighbor ferromagnetic exchange $J_1$ and the next-nearest-neighbor antiferromagnetic exchange $J_2$. Lately, it has become controversial whether $J_2$ is substantially stronger in magnitude than $J_1$. We resolved this controversy by determining the crystal structure of $LiCuVO_4$ at 1.6 K from neutron diffraction, analyzing the magnetic susceptibility of $LiCuVO_4$, and determining the spin exchange constants of $LiCuVO_4$ from density functional calculations.

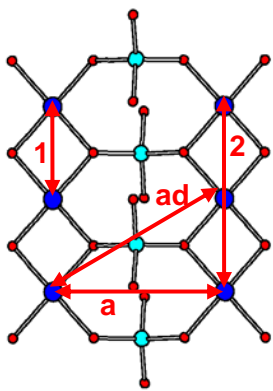
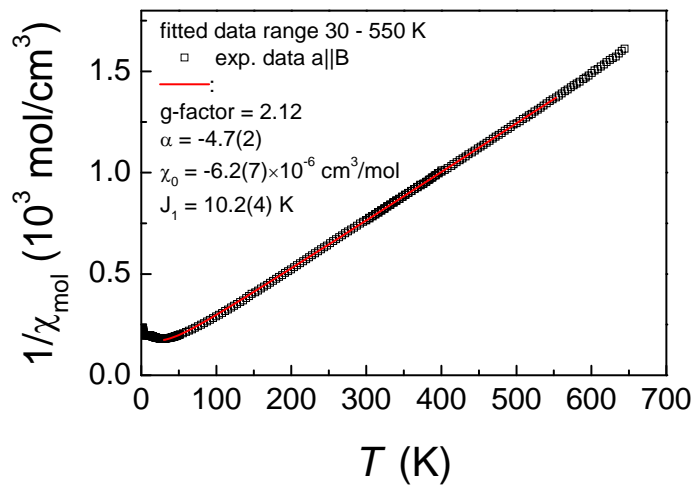